\documentclass[aps,prd,preprint,groupedaddress,preprintnumbers,floatfix]{revtex4}
\usepackage{epsf}
\usepackage{epsfig}
\usepackage{graphics}
\usepackage{subfigure}
\usepackage{xspace}

\newcommand{\ra} {$\rightarrow$}

\newcommand{\be}{\begin{equation}}
\newcommand{\ee}{\end{equation}}

\newcommand{\bea}{\begin{eqnarray}}
\newcommand{\eea}{\end{eqnarray}}
\newcommand{\beanon}{\begin{eqnarray*}}
\newcommand{\eeanon}{\end{eqnarray*}}
\newcommand{\ba}{\begin{array}}
\newcommand{\ea}{\end{array}}
\newcommand{\bd}{\begin{description}}
\newcommand{\ed}{\end{description}}
\newcommand{\bi}{\begin{itemize}}
\newcommand{\ei}{\end{itemize}}
\newcommand{\ben}{\begin{enumerate}}
\newcommand{\een}{\end{enumerate}}
\newcommand{\bc}{\begin{center}}
\newcommand{\ec}{\end{center}}

\newcommand{\toptop}{\mbox{${\mathrm t} \bar {\mathrm t}$}\xspace}

\newcommand{\VVL}{\mbox{${\mathrm V_L}{\mathrm V_L}$}\xspace}

\newcommand{\WW}{\mbox{${\mathrm W}{\mathrm W}$}\xspace}
\newcommand{\ZZ}{\mbox{${\mathrm Z}{\mathrm Z}$}\xspace}
\newcommand{\VV}{\mbox{${\mathrm V}{\mathrm V}$}\xspace}
\newcommand{\ZW}{\mbox{${\mathrm Z}{\mathrm W}$}\xspace}

\newcommand{\W}{\mbox{${\mathrm W}$}\xspace}

\newcommand{\VL}{\mbox{${\mathrm V_L}$}\xspace}
\newcommand{\Z}{\mbox{${\mathrm Z}$}\xspace}
\newcommand{\pt}{\mbox{${\mathrm p_T}$}\xspace}

\newcommand{\eqn}[1]{Eq.(\ref{#1})}

\newcommand{\tbn}[1]{Tab.~\ref{#1}}

\newcommand{\fig}[1]{Fig.~\ref{#1}}
\newcommand{\figs}[2]{Figs.~\ref{#1}--\ref{#2}}

\newcommand{\sect}[1]{Sect.~\ref{#1}}
\newcommand{\subsect}[1]{Sub-Sect.~\ref{#1}}

\newcommand{\Ref}[1]{Ref.\cite{#1}}

\renewcommand{\O}{{\mathcal O}}

\newcommand{\Phase}{{\tt PHASE}\xspace}

\newcommand{\Phantom}{{\tt PHANTOM}\xspace}

\def\pl #1 #2 #3 {{\it Phys.~Lett.} {\bf#1} (#2) #3}   
\def\np #1 #2 #3 {{\it Nucl.~Phys.} {\bf#1} (#2) #3}
\def\zp #1 #2 #3 {{\it Z.~Phys.} {\bf#1} (#2) #3}
\def\pr #1 #2 #3 {{\it Phys.~Rev.} {\bf#1} (#2) #3}
\def\prep #1 #2 #3 {{\it Phys.~Rep.} {\bf#1} (#2) #3}
\def\prl #1 #2 #3 {{\it Phys.~Rev.~Lett.} {\bf#1} (#2) #3}
\def\intj #1 #2 #3 {{\it Int. J. Mod. Phys.} {\bf#1} (#2) #3}
\def\mpl #1 #2 #3 {{\it Mod.~Phys.~Lett.} {\bf#1} (#2) #3}
\def\rmp #1 #2 #3 {{\it Rev. Mod. Phys.} {\bf#1} (#2) #3}
\def\cpc #1 #2 #3 {{\it Comp. Phys. Commun.} {\bf#1} (#2) #3}
\def\epj #1 #2 #3 {{\it Eur. Phys. J.} {\bf#1} (#2) #3}
\def\jhep #1 #2 #3 {{\it JHEP} {\bf#1} (#2) #3}

\begin{document}

\title{Boson Fusion and Higgs production at the LHC in six fermion final states
with one charged lepton pair.}

\author{Elena Accomando}
\email{accomand@to.infn.it}

\author{Alessandro Ballestrero}
\email{ballestr@to.infn.it}
\author{Aissa Belhouari}
\email{belhouar@to.infn.it}
\author{Ezio Maina}
\email{maina@to.infn.it}

\affiliation{
          INFN, Sezione di Torino and
  Dipartimento di Fisica Teorica, Universit\`a di Torino\\
  Via Giuria 1, 10125 Torino, Italy}

\thanks{
E.A. is supported by the Italian Ministero dell'Istruzione, 
dell'Universit\`a e della Ricerca (MIUR) under contract Decreto MIUR 
26-01-2001 N.13 ``Incentivazione alla mobilit\`a di studiosi stranieri ed 
italiani residenti all'estero''.\\
Work supported by MIUR under contract 2004021808\_009.
}

\begin{abstract}
Boson boson scattering and Higgs production in
boson boson fusion will be actively investigated at the LHC.
We have performed a parton level study of all processes of the type 
$q_1 q_2 \rightarrow q_3 q_4 q_5 q_6 l^+l^-$ using for
the first time a full fledged six fermion Monte Carlo event generator
which employs exact matrix elements at $\O(\alpha_{em}^6)$.
We have examined Higgs production in vector boson fusion followed by the decay
chain $H\rightarrow ZZ\rightarrow l^+l^-jj$,
including exactly all electroweak irreducible backgrounds.
In the high mass region we have compared the case of a relatively light Higgs
with the no-Higgs results. The integrated cross section for the latter case is
more than twice that in the former for a minimum
invariant mass of the $ZV$ pair of about 800 GeV. We find, in a preliminary
analysis at parton level that, summing up the muon and the
electron channels, about 25 events are expected in the light Higgs case
for L=100 $fb^{-1}$.
\end{abstract}

\preprint{DFTT 04/2006}
\maketitle


\section{Introduction}
\label{sec:intro}
The Standard Model (SM) provides the simplest and most economical
explanation of Electro--Weak Symmetry Breaking (EWSB). Detailed reviews and extensive
bibliographies can be found in 
Refs.\cite{HiggsLHC,djouadi-rev1,ATLAS-TDR,Houches2003}.
The only missing ingredient is the Higgs boson.
The fit of the SM to precision EW data currently gives an upper limit on the
Higgs mass of about 200 GeV \cite{lepewwg} while direct searches have
established a 95\% CL lower bound M(H)$>$114.4 GeV \cite{lepHiggs}.

In the SM the Higgs is essential to the
renormalizability of the theory and is also crucial to 
ensure that perturbative unitarity bounds are not violated in high energy
reactions. Scattering processes between longitudinally polarized vector bosons
(\VL) are particularly sensitive in this regard. Without a Higgs the \VL's
interact strongly at high energy,
violating perturbative unitarity at about one TeV \cite{reviews}.
If, on the contrary, a relatively light Higgs exists then
they are weakly coupled at all energies.
In the strong scattering case
one is led to expect the presence of resonances in 
\VVL interactions. Unfortunately the mass, spin and even number of these
resonances are not uniquely determined \cite{unitarization,butterworth02}.
If a Higgs particle is discovered it will nonetheless
be necessary to verify that
indeed longitudinally polarized vector bosons are weakly coupled at high energy
by studying boson boson scattering in full detail.

At the LHC no beam of on shell EW bosons will be available. Incoming quarks will
emit spacelike virtual bosons which will then scatter among themselves and 
finally decay.
These processes have been scrutinized since
a long time, going from the pioneering works in \cite{history1,history2}, which
address
boson boson scattering on a general ground, to the more recent papers in
\cite{history3,history4} focused on the extraction of signals of vector boson
scattering at the LHC. 
All previous studies of boson boson scattering at high energy hadron colliders,
with the exception, to our knowledge, of \cite{Accomando:2005hz}
and \cite{Eboli:2006wa},
have resorted to some approximation, either the Equivalent Vector Boson
Approximation (EVBA) \cite{EVBA},
or a production times decay approach, supplementing a calculation of 
\begin{equation}
\label{2f2b}
q_1 q_2 \rightarrow q_3 q_4 V_1 V_2  
\end{equation}
processes with the, typically on shell, decay of the two vector bosons.
There are however issues that cannot be tackled without a full six fermion
calculation like exact spin correlations between the decays of different heavy
particles, the effect of the non resonant background, the relevance of the
offshellness of boson decays, the question of interferences between
different
subamplitudes. Without a complete calculation it will be impossible to
determine the accuracy of approximate results. In Ref.\cite{Accomando:2005hz}
this
issue was discussed at lenghth, showing differences of the order of 10--20\% in
some important regions of phase space.
The reliability of the EVBA approximation in the context of vector boson
scattering has been critically examined in \cite{EVBAvsComplete}.

Recently \Phase a full fledged six fermion Monte Carlo has become available
\cite{ref:Phase}. It describes at 
{$\O(\alpha_{em}^6)$}, using exact tree level matrix elements,
all processes of the form
$PP \rightarrow q_1 q_2 \rightarrow q_3 q_4 q_5 q_6 l \nu$  
(where $q_i$ stands for a generic (anti)quark) which can take place at the LHC
\cite{ACAT-QFTHEP,Accomando:2005hz}.
The range of interesting reactions is however much wider. Processes in which
both vector bosons decay leptonically have been extensively studied both for
Higgs detection and for boson boson scattering and top physics. Besides, 
in order to obtain a full coverage of all semileptonic processes it is necessary
to include all reactions with a charged lepton pair in the final state. 
This has required the calculation of additional amplitudes and an extensive
improvement of the routines which pilot the integration and the generation of
unweighted events. The result is a new code called \Phantom 
\cite{Phantom} which, at present,     
includes all processes with six fermions in the final state at
{$\O(\alpha_{em}^6)$}
\begin{equation}
\label{6f}
PP \rightarrow q_1 q_2 \rightarrow f_1 f_2 f_3 f_4 f_5 f_6.  
\end{equation}
\noindent
The accuracy of tree level calculations can be sensibly improved.
\Phase is being continued also in this direction.
In particular, electroweak corrections have proved to be sizable expecially for
processes involving the Higgs boson, see for instance \Ref{ewcor}.  A new code
{\tt PHAST\_NLO} \cite{Accomando:PHAST_NLO} will address
$O(\alpha_{em})$ electroweak radiative effects in six fermion physics.
Both \Phantom and {\tt PHAST\_NLO} are based on the methods of
Refs. \cite{method,phact} and adopt the 
iterative-adaptive multichannel strategy developed in \cite{ref:Phase}.
In the following we present results obtained with \Phantom.

\Phantom is an example of {\it dedicated} event generator which describes
a predefined set of reactions striving for maximum speed and efficiency.
Other recent examples of {\it dedicated} programs for LHC physics are 
{\tt Alpgen} \cite{Mangano:2002ea} and {\tt Toprex} \cite{Slabospitsky:2002ag}. 
The complementary approach is 
given by {\it multi-purpose} programs for the automatic generation of any 
user-specified parton level process. The following codes for multi-parton 
production are available: {\tt Amegic-Sherpa} \cite{Amegic}, {\tt CompHEP} 
\cite{Comphep}, {\tt Grace-Gr@ppa} \cite{Grace}, {\tt Madevent} 
\cite{Madgraph}, {\tt Phegas \& Helac} \cite{Phegas}, 
{\tt O'Mega $\&$ Whizard} \cite{Omega}.

Recently the $\O(\alpha_{em}^6\alpha_s)$ QCD corrections to 
$q_1 q_2 \rightarrow q_3 q_4 e^+ \nu_e  \mu^-\overline{\nu}_\mu$ a 
 have been computed in \cite{Jager:2006zc}
and those to 
$q_1 q_2 \rightarrow q_3 q_4 e^+ e^-  \mu^+\mu^-$ 
and 
$q_1 q_2 \rightarrow q_3 q_4 e^+ e^-  \nu_\mu \overline{\nu}_\mu$ 
in  
\cite{Jager:2006cp}.
They turn out to be modest, changing the total cross sections by less than 10\%.
The smallness of QCD corrections is related to color conservation which forbids gluon
exchange between quark lines, up to higly suppressed contributions generated by the exchange
of identical quarks. The same applies to 
QCD corrections to $q_1 q_2 \rightarrow q_3 q_4 q_5 q_6 l^+ l^-$ which however do include
a larger number of terms which could result in larger corrections.
It should be pointed out that QCD effects are not limited to NLO corrections.
In the context of vector boson pair production
$ PP \rightarrow WW$, gluon initiated
$\O(\alpha_s^2)$ processes $gg \rightarrow WW$, proceeding through a quark
loop, have been shown to be a sizable correction to the tree level reactions
$q\bar{q} \rightarrow WW$ when realistic Higgs search cuts
are imposed \cite{Binoth:2005ua}. On the other hand QCD $WWjj$ production 
represents only a small background to Higgs searches in Vector Boson
Fusion \cite{Atlas_HinWW}.

\begin{figure}
\begin{center}
\mbox{\epsfig{file=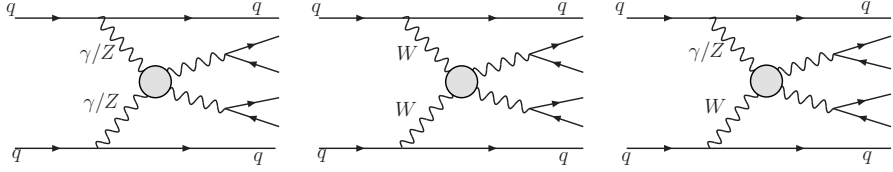,width=12.cm}}
\caption{Vector boson fusion processes.}
\label{VV-diag}
\end{center}
\end{figure}

Since in addition to \VV scattering many other subprocesses are in general
present in the full set of diagrams, as partially 
shown in \figs{VV-diag}{higgs-diag}, it is not a trivial task to separate
boson boson scattering from the EW irreducible background.
In practice one has to deal also with other types of
background to which
QCD interactions contribute, but which however do not include any boson boson
scattering term. We will refer to these processes as QCD
background even though in general they will be a mixture of QCD and EW
interactions.
In this paper we are neglecting QCD backgrounds.
It is clear that obtaining a good signal over EW irreducible background ratio
is a prerequisite to any attempt at dealing with the QCD one.

We are aware that much is still needed to obtain a thoroughly realistic
assessment of the observability of these processes. Only a complete study
including $4jl^+l^-$ at $\O(\alpha_{em}^4\alpha_s^2)$ and
$\O(\alpha_{em}^2\alpha_s^4)$ together with full detector simulation will be
able to say the final word. In the meantime it is important that the tools
available for simulation are sharpened as much as possible and that the
viability of such analyses is demonstrated at $\O(\alpha_{em}^6)$ which includes
all signal contributions. In our opinion, the results presented in the
following and the event generator used to produce them represent a step forward
in this direction.

\begin{figure}
\begin{center}
\mbox{\epsfig{file=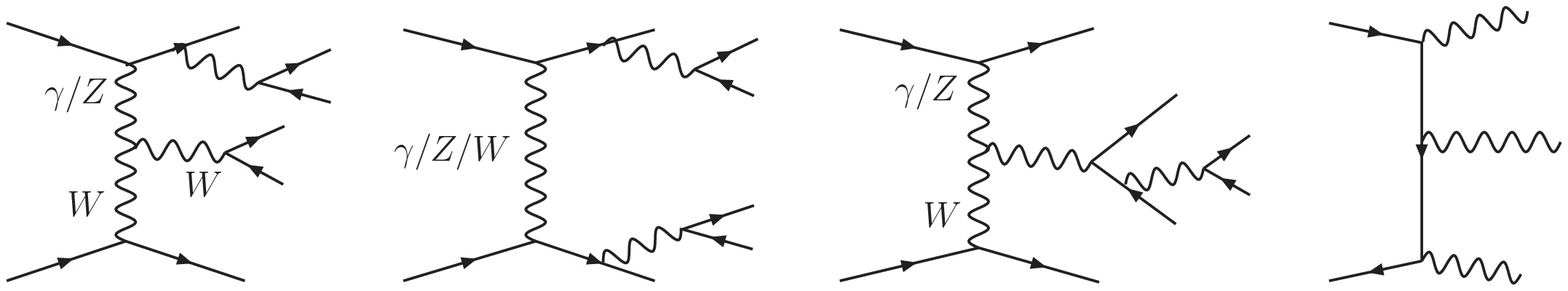,width=14.cm}}
\caption{ Examples of EW irreducible background to vector boson scattering
processes.}
\label{nonreso-diag}
\end{center}
\end{figure}

\begin{figure}
\begin{center}
\mbox{\epsfig{file=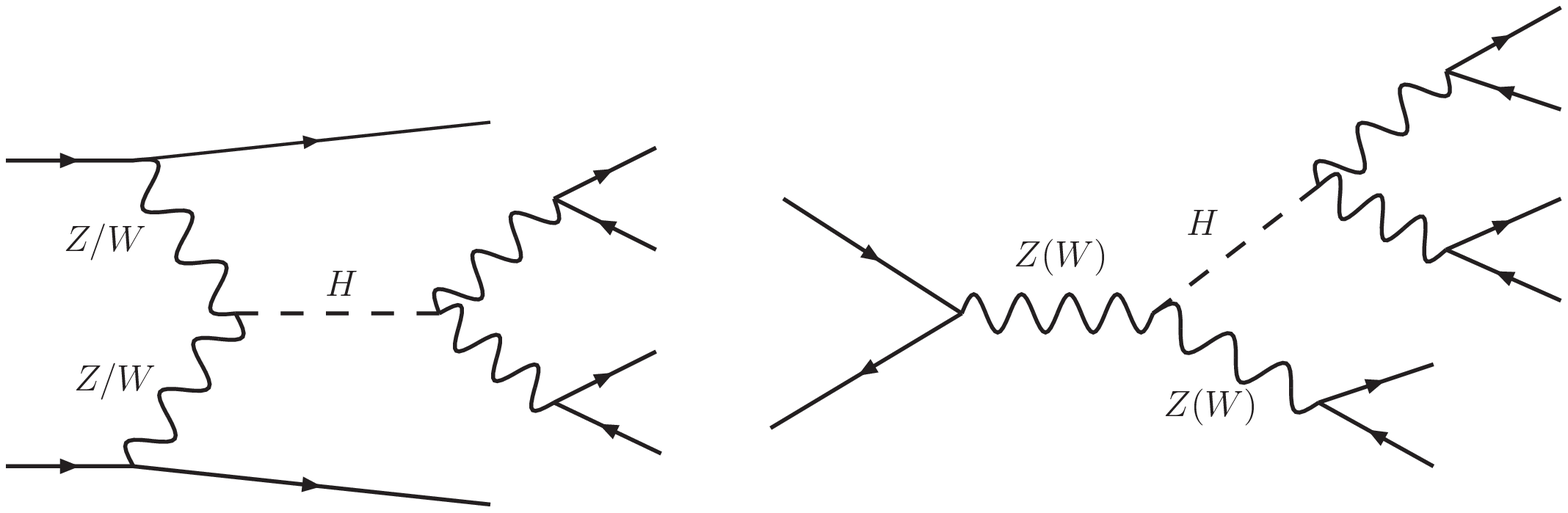,width=12.cm}}
\caption{Higgs boson production via vector boson fusion
  and Higgsstrahlung.}
\label{higgs-diag}
\end{center}
\end{figure}

\section{Classification and calculation} 
\label{sec:Calculation}

\begin{table}[htbp]
\begin{center}
\begin{tabular}{|c|c|c|c|c|c|}
\hline
Group& Type & diagrams & Group& Type & diagrams \\
\hline
\vspace{-0.2cm}
  & & & & & \\
$u\bar{u}\,c\bar{c}\,b\bar{b}\,l^-l^+$& \rule{6mm}{0mm}4Z  \rule{6mm}{0mm}& 615   & $d\bar{d}\,s\bar{s}\,b\bar{b}\,l^-l^+$& 4Z & 615  \\  
\hline
\vspace{-0.2cm}
  & & & & & \\
$u\bar{u}\,u\bar{u}\,b\bar{b}\,l^-l^+$& 4Z & 1230  & $u\bar{u}\,u\bar{u}\,u\bar{u}\,l^-l^+$& 4Z & 3474  \\
\hline
\vspace{-0.2cm}
  & & & & & \\
$u\bar{u}\,u\bar{u}\,s\bar{s}\,l^-l^+$& 4Z & 1158  & $u\bar{u}\,u\bar{u}\,c\bar{c}\,l^-l^+$& 4Z & 1158  \\
\hline
\vspace{-0.2cm}
  & & & & & \\ 
$u\bar{u}\,b\bar{b}\,b\bar{b}\,l^-l^+$& 4Z & 1606  & $u\bar{u}\,c\bar{c}\,d\bar{d}\,l^-l^+$&4Z+2W2Z&821 \\
\hline
\vspace{-0.2cm}
  & & & & & \\
$u\bar{u}\,s\bar{s}\,s\bar{s}\,l^-l^+$& 4Z & 1158  & $u\bar{u}\,u\bar{u}\,d\bar{d}\,l^-l^+$&4Z+2W2Z&2126\\
\hline
\vspace{-0.2cm}
  & & & & & \\
$u\bar{u}\,s\bar{s}\,b\bar{b}\,l^-l^+$& 4Z & 615   & $u\bar{u}\,d\bar{d}\,b\bar{b}\,l^-l^+$& 4Z+2W2Z&880\\
\hline
\vspace{-0.2cm}
  & & & & & \\
$d\bar{d}\,d\bar{d}\,d\bar{d}\,l^-l^+$& 4Z & 3474  & $u\bar{u}\,d\bar{d}\,s\bar{s}\,l^-l^+$&4Z+2W2Z& 821\\
\hline
\vspace{-0.2cm}
  & & & & & \\
$b\bar{b}\,b\bar{b}\,b\bar{b}\,l^-l^+$& 4Z & 7506  & $u\bar{u}\,d\bar{d}\,d\bar{d}\,l^-l^+$&4Z+2W2Z&2126\\
\hline
\vspace{-0.2cm}
  & & & & & \\
$d\bar{d}\,d\bar{d}\,s\bar{s}\,l^-l^+$& 4Z & 1158  & $u\bar{d}\,s\bar{c}\,u\bar{u}\,l^-l^+$& 2W2Z & 484\\
\hline
\vspace{-0.2cm}
  & & & & & \\
$d\bar{d}\,d\bar{d}\,b\bar{b}\,l^-l^+$& 4Z & 1230  & $u\bar{d}\,s\bar{c}\,d\bar{d}\,l^-l^+$& 2W2Z & 484\\
\hline
\vspace{-0.2cm}
  & & & & & \\
$d\bar{d}\,b\bar{b}\,b\bar{b}\,l^-l^+$& 4Z & 1606  & $u\bar{d}\,s\bar{c}\,b\bar{b}\,l^-l^+$& 2W2Z & 265\\
\hline
  
\end{tabular}
\end{center}
\caption{Classification of 
$p p\rightarrow q q^\prime \rightarrow 4 q+l^+l^-$ processes.
The first column shows the group list, the second the process type as discussed
in the text,
the third the corresponding number of diagrams.} 
\label{4z_proc}
\end{table}

For a complete analysis one needs to include all processes which
contribute  to final states with one charged lepton pair. Taking into 
account one lepton type, charge conjugation
 and the symmetry between the first and second quark families, 
the number of reactions can be reduced to 135.  A given reaction, its 
charge-conjugate, the ones related by family exchange and those obtained with
the product of the two transformations can be indeed 
described by the same matrix element; they differ by the convolution
with Parton Distribution Functions (PDF).
All processes which share 
the same total particle content, with all eight partons taken to be outgoing, 
can be described by a single master amplitude.  
As a consequence, all reactions  can be classified 
into 22 groups which are enumerated in \tbn{4z_proc}. 
By selecting two initial quarks in each particle group, one obtains all
possible processes.  

The calculation can be further simplified examining more closely the full set
of Feynman diagrams. In some processes, fermions can be paired only into neutral
currents (4Z), while in other cases they can form two charged and two neutral
currents (2Z2W). Mixed processes are described by a combination of
the two sets (2Z2W+4Z). 

\begin{figure}[t!b]
\begin{center}
  \unitlength 1cm
  \begin{picture}(10,4.5) (0,0)
    \put(-1.3,0.){ \scalebox{.40}{\includegraphics*{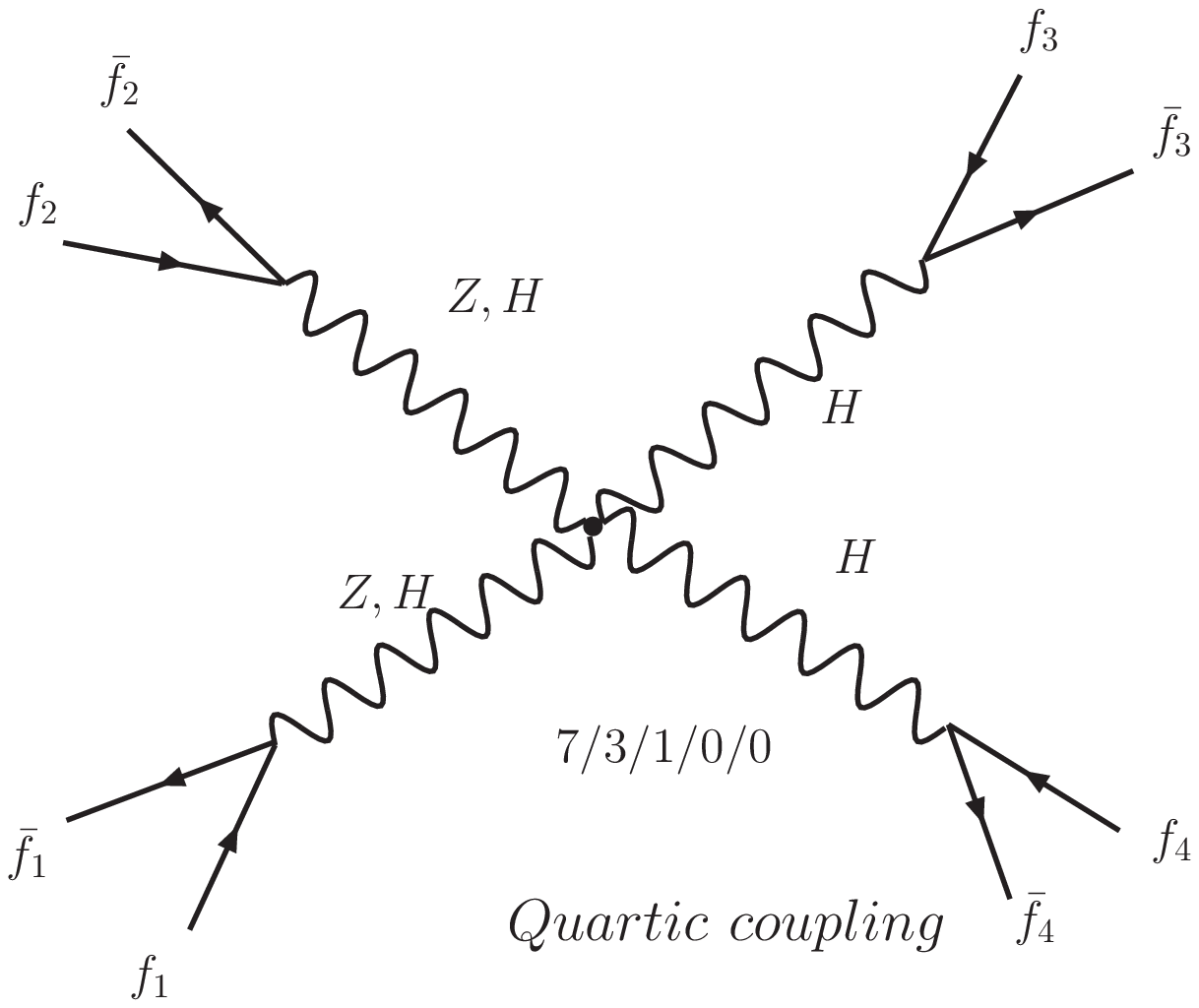}} }
    \put(4.2,0.){ \scalebox{.45}{\includegraphics*{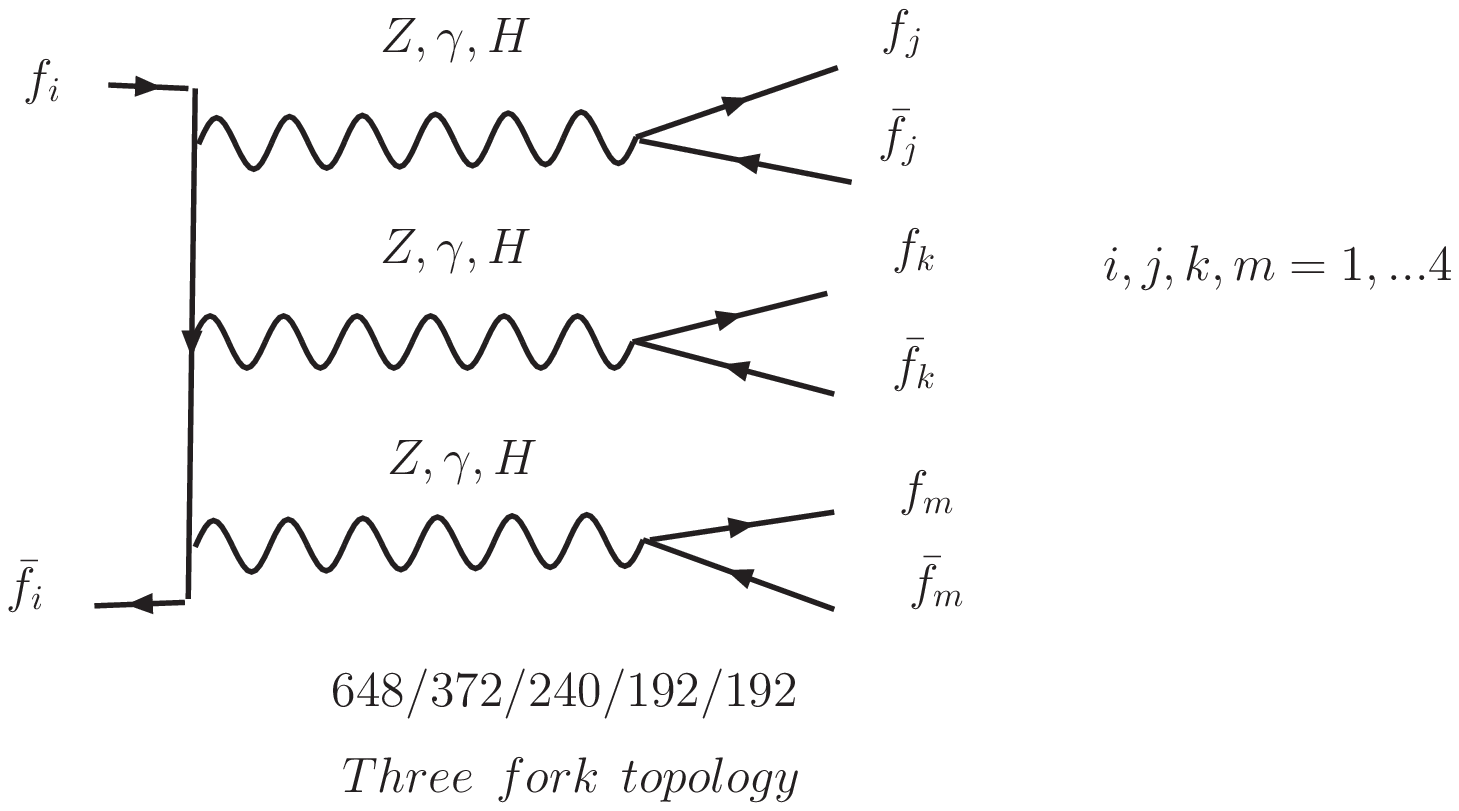}} }
    \put(2.7,-3.3){ \scalebox{.5}{\includegraphics*{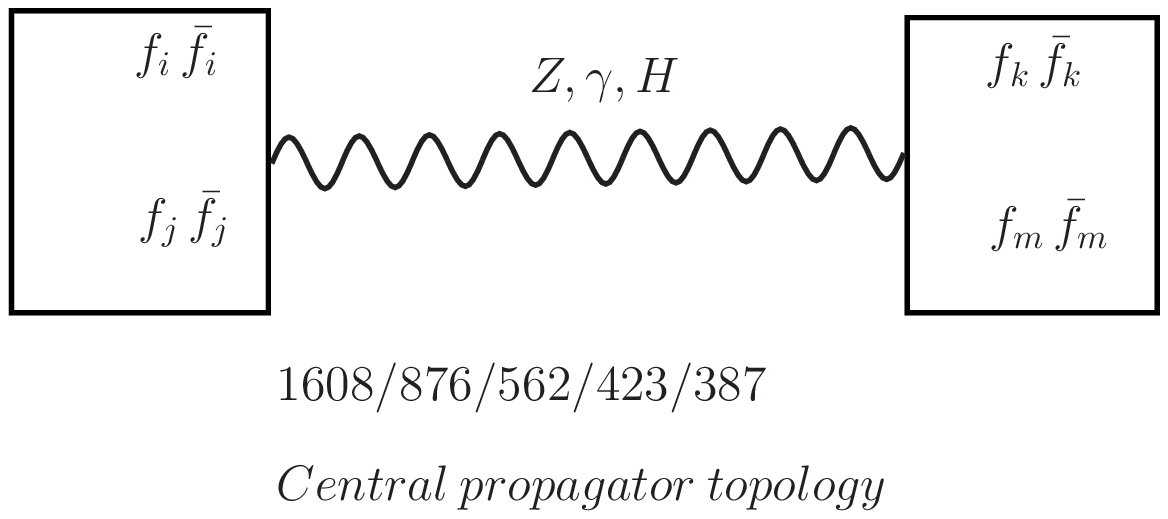}} }
 \end{picture}
    \vspace{3cm}
    \caption{Fundamental topologies associated with 4Z processes.
     Total number of diagrams 2263/1251/803/615/579 }
    \label{topologies}
  \end{center}
\end{figure}

The 4Z amplitude was not previously available.
The three basic topologies in which the Feynman diagrams appearing in
the 4Z amplitude can be classified are shown in \fig{topologies}.
The numbers N4/N3/N2/N1/N0  under each topology indicate the number of
Feynman diagrams described by this topology 
if 4/3/2/1/0 fermion pairs are taken to be massive, namely to have non zero
interaction with the Higgs boson. The flavour of all four fermion pairs are
assumed to be different.
The complete set of diagrams is eventually generated by 
exchange of identical particles. 

\begin{figure}[hbt]
  \begin{center}
    \includegraphics[width=12cm]{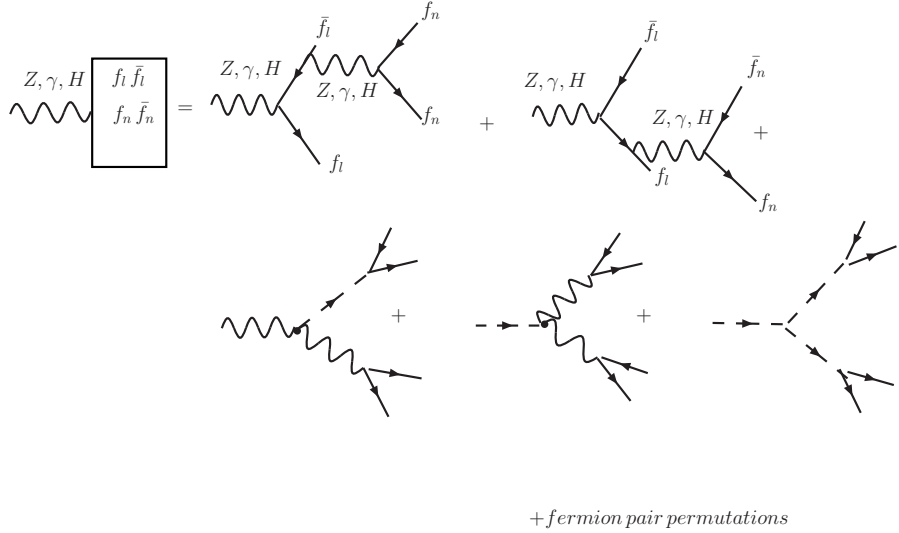}
\caption{Decays of the off shell neutral bosons $Z,\: \gamma, \: H $}
 \label{square}
  \end{center}
\end{figure}

Each rectangle on both sides of the central boson topology is the sum of
several subdiagrams representing all possible decays of an off
shell $Z,\: \gamma$ or Higgs boson to four outgoing particles as shown in
\fig{square}.
These sets of subdiagrams are evaluated only once, with a substantial
efficiency gain, and then combined together in the end.  

\section{Physical processes}
\label{PhysSub}

Boson boson scattering and Higgs production in boson boson fusion
produce intermediate states with two bosons and two quarks as shown
in \fig{VV-diag}. 
In this study we have only considered final states in which one \Z boson 
decays leptonically to $\mu^+\mu^-$
and the other (either \Z or \W) hadronically.
If both bosons decay hadronically the signal cannot
be distinguished from the QCD background 
whose cross section is much larger. 
Final states where both vectors decay leptonically 
have a smaller rate and have been left for future studies.

\begin{figure}
\begin{center}
\mbox{
{\epsfig{file=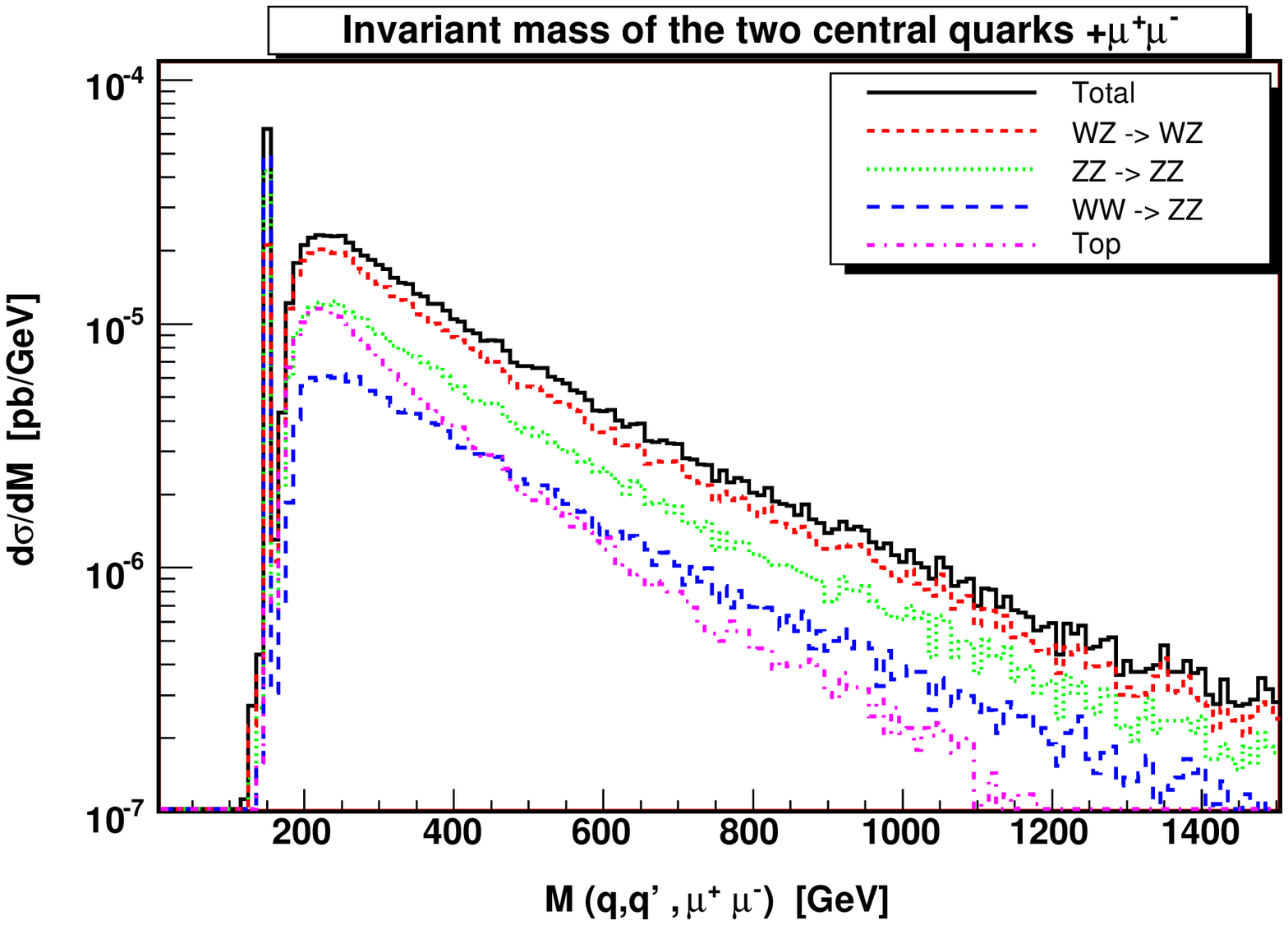,width=13cm}}
}
\mbox{
{\epsfig{file=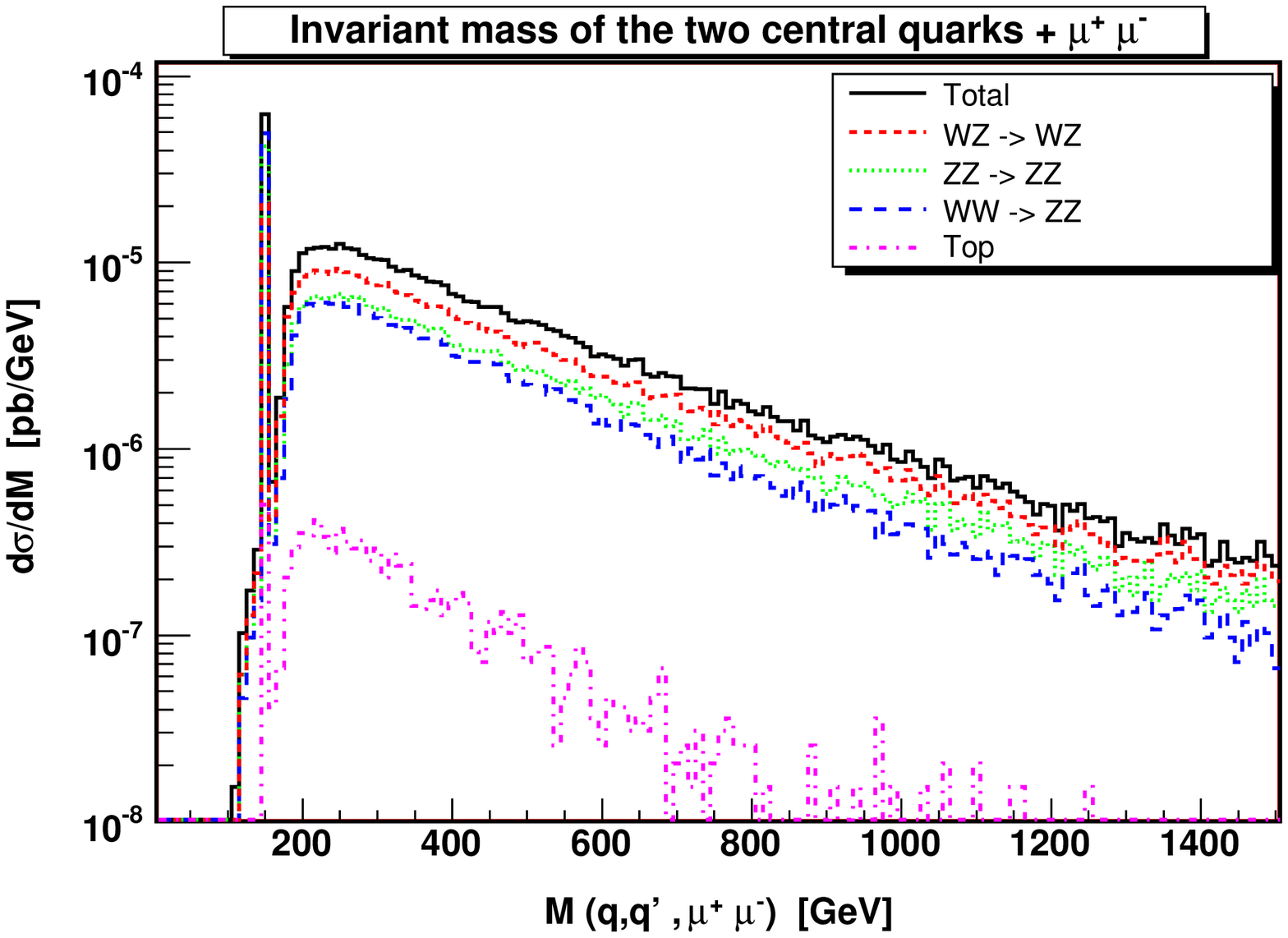,width=13cm}}
} 
\caption{Invariant mass distribution of the 
two charged leptons and the two most central quarks, for different sets
of processes. The upper plot includes the set of cuts described in the text.
In the lower plot a further cut for vetoing top production is applied.} 
\label{sub-proc}
\end{center}
\end{figure}

\begin{table}[bth]
\begin{center}
\begin{tabular}{|c|}
\hline
      \pt(lepton)$> 20$ GeV \\
\hline
      $\vert\eta$(lepton)$\vert<3$ \\
\hline
      E(quark)$>30$ GeV \\
\hline
      \pt(quark)$>20$ GeV \\
\hline
      $\vert\eta$(quark)$\vert<5$ \\
\hline
      M($l^+l^-$)$>20$ GeV \\
\hline
      M(qq)$>60$ GeV \\
\hline
      $\Delta \eta$(tag-quarks)$> 3.8$ \\
\hline
\end{tabular}
\caption{Standard acceptance cuts applied in all results. Any pair of colored
fermions must have mass larger than 60 GeV. $\Delta \eta$(tag-quarks) is the
absolute value of the difference in pseudorapidity between the two tag quarks.} 
\label{standard-cuts}
\end{center}
\end{table}
 
A number of event samples representative of all possible processes of the form
$PP \rightarrow q_1 q_2 \rightarrow q_3 q_4 q_5 q_6 l^+ l^-$  have been
produced with \Phantom. In order to comply with  typical
acceptance and trigger requirements, the cuts in \tbn{standard-cuts} have
been applied. The acceptance cuts are standard. The wide separation in $\eta$
between the two tag quarks is a well established technique for separating the
scattering signal from the background \cite{history2,history3,history4}.
We have imposed a minimum invariant mass cut of 60 GeV on jet
pairs rather than requiring a minimum $\Delta R$ separation. It is well known
that at large $p_T$ the two jets from the hadronic decay of a \W or \Z boson
tend to coalesce. This issue has been discussed, for instance,
at length in the ATLAS TDR \cite{ATLAS-TDR}
in connection with the observability of a
heavy Higgs decaying to a \WW pair, where a number of jet finding schemes have
been studied.We defer to \subsect{sec:highmass}
a discussion of the effects of a separation in $\Delta R$ for the class of
processes under consideration.
It should be kept in mind that selection procedures can be optimized in
different ways for different
analyses and that such optimization has to be performed at hadron
level rather than at the more primitive parton stage we are discussing here.

In most of this paper, since we are mainly concerned with extracting signals of
boson boson scattering from the irreducible background, we select events using
flavour information, which will be unavailable in actual analyses, for the
identification of vector bosons and top quarks. In \subsect{sec:highmass} we
show that our results are not substantially modified if one adopts a more
realistic procedure entirely based on invariant mass cuts.

We have used the CTEQ5L \cite{CTEQ5} PDF set with scale
\be
Q^2 = M_W^2 + \frac{1}{6}\,\sum_{i=1}^6 p_{Ti}^2.
\label{scale}
\ee
where $p_{Ti}$ denotes the transverse momentum of the $i$--th final state
particle.
 
Many subprocesses 
(i.e. \ZW \ra \ZW, \WW \ra \ZZ, \ZZ \ra \ZZ, $\mathrm qb$ \ra $\mathrm qtV$)
will in general contribute to a specific six fermion reaction. 
\toptop processes will not contribute to the $4ql^+l^-$ channels
but single top production with an additional neutral boson emission will be
present.

It is  impossible to separate and compute individually the cross 
section due to a single subprocess, since there are large interference effects
between the different contributions.
We can however select all complete 2\ra6 processes which include a specific set
of subdiagrams.
For instance, \ZW\ra\ZW with on shell bosons is described by four Feynman diagrams.
These same diagrams, with the two incoming external vector bosons connected to
the initial fermion lines and the two final ones connected to their decay 
products, constitute the  \ZW\ra\ZW set of 2\ra6 diagrams.
Several sets can contribute to a single
process and therefore the same process can appear in different groups;
for example $ud \rightarrow ud \mu^+\mu^- b \bar{b}$ will be included in both
the \WW \ra \ZZ and \ZZ \ra \ZZ groups.
As a consequence of this multiple counting, the total cross section
is smaller than the sum of the cross sections for the various groups.
The upper part of \fig{sub-proc} shows the invariant mass distribution
of the two most central quarks (when ordered in pseudorapidity
$\eta$) and of the two leptons for all reactions which contain the
different subprocesses
as well as the distribution for the complete set of processes.
We assumed M(H)=150 GeV.
Notice that the Higgs peak is present in the \ZW\ra\ZW curve. This is due to
processes, like for instance $u\bar{d} \rightarrow u\bar{d} d\bar{d} \mu^+\mu^-$
which in addition to the \ZW\ra\ZW set of diagrams
(e.g. $u\bar{d}  \rightarrow d\bar{d} W Z$)
include also
diagrams describing Higgs production in $s$--channel
(e.g. $u\bar{d} \rightarrow  u\bar{d} Z Z$).

The group comprising top diagrams has a large
cross section. 
The lower part of \fig{sub-proc} shows
the same distributions after top subtraction. 
Top candidates are identified requiring a b-quark and two other quarks
in the final state of the right flavour combination to
be produced in a W decay, with a total  invariant mass between 160 and 190
GeV.

If no Higgs is present, all SM scattering processes between on shell weak
vector bosons grow linearly with the center of mass energy squared,
with the exception
of \ZZ \ra \ZZ which in this case  is zero. 
This behaviour is in agreement with the low energy theorem (LET)
\cite{LowEnergyTheorem}.
The lower part of \fig{sub-proc} shows that the \ZZ \ra \ZZ component
is relatively small compared
with the total distribution and as a consequence does not represent too serious
a background to searches for new physics signals. 
The group including the \ZW\ra\ZW set of diagrams gives the largest
contribution.

In \tbn{total-sigma} we present the total cross section for
$p p\rightarrow q q^\prime \rightarrow 4 q+\mu^+mu^-$
with the standard acceptance cuts in \tbn{standard-cuts} for different Higgs
masses.In \tbn{total-sigma-scale} we show the scale dependence of the total
cross section for two Higgs scenarios, adopting the usual recipe of varying the
scale by a factor of two in either direction. This leads to a 4\% difference
with respect to the central value.  

\begin{table}[tbh]
\begin{center}
\begin{tabular}{|c|c|}
\hline
$M_{H}$ & Cross section (fb) \\
\hline
150 GeV & 7.33$\pm$0.02 \\
\hline
200 GeV & 13.67$\pm$0.03 \\
\hline
500 GeV & 9.89$\pm$0.02 \\
\hline
NoHiggs & 7.34$\pm$0.02 \\
\hline
\end{tabular}
\caption{Total cross section for $p p\rightarrow q q^\prime \rightarrow 
4 q+\mu^+mu^-$
with the standard acceptance cuts in \tbn{standard-cuts} for different
Higgs masses.} 
\label{total-sigma}
\end{center}
\end{table}

\begin{table}[tbh]
\begin{center}
\begin{tabular}{|c|c|c|c|}
\hline
$M_{H}$ & Q/2 & Q & 2Q\\
\hline
200 GeV & 14.21 $\pm$0.02 & 13.67$\pm$0.03 & 13.11 $\pm$0.02 \\
\hline
NoHiggs & 7.57 $\pm$0.02  & 7.34$\pm$0.02 & 7.01 $\pm$0.02 \\
\hline
\end{tabular}
\caption{Scale dependence of the total cross section for
$p p\rightarrow q q^\prime \rightarrow 
4 q+\mu^+mu^-$
with the standard acceptance cuts in \tbn{standard-cuts} for two Higgs
scenarios. $Q$ is defined in \eqn{scale}} 
\label{total-sigma-scale}
\end{center}
\end{table}
\subsection{Higgs production and its complete EW background in \Phantom}

 \Phantom is capable of simulating Higgs production
in \VV fusion together with all its EW irreducible background for all channels
and
for any Higgs mass and may be particularly useful in the intermediate
mass range, below the \WW threshold, where the production times decay approach
cannot be used.
Its improved treatment of the EW sector needs to be complemented by an accurate
description of QCD dominated backgrounds and of the effects of QCD NLO
corrections.

Higgs production in \VV fusion followed by Higgs decay to \WW or \ZZ is 
the second most abundant production channel over almost the full range of 
Higgs masses which will be explored at the LHC. It is regarded
as the channel with the highest statistical significance for an intermediate
mass Higgs \cite{Atlas_HinWW,CMS:Higgs1}.
Previous analyses have focused 
mainly
on the \WW channel.
For an intermediate mass Higgs the dilepton final state
$H \rightarrow WW^{(\ast)} \rightarrow l\nu l\nu$ is slightly favoured with
respect to the $H \rightarrow WW^{(\ast)} \rightarrow l\nu jj$ channel because
of the $W + nj$ background which affects the latter. In the first case the main
background comes from $t \bar t$ production followed in importance
by EW $WWjj$ production which is estimated to be about 10\% of the signal.
In the second case the main background comes from $W + nj$ followed by
$t \bar t$ and EW $WWjj$ production.
QCD $WWjj$ production can be reduced to be of the same order of magnitude as
the EW contribution using acceptance cuts and 
can be rendered essentially negligible by a central jet veto which is not so
effective in the EW case \cite{Atlas_HinWW,history4}.

The production channel 
$qq\rightarrow qqH,\ H\rightarrow ZZ \rightarrow l^+l^-jj$ has been examined in
\cite{Atlas_HinVBF} while the channels
$qq\rightarrow qqH,\ H\rightarrow ZZ \rightarrow l^+l^-\nu\overline{\nu}$ and
$ ZZ \rightarrow l^+l^-l^+l^-$ have been considered in \cite{CMS:Higgs1}.
The $l^+l^-jj$ and $l^+l^-l^+l^-$ channels are particularly interesting 
because they allow a direct reconstruction of the Higgs mass
which in $l\nu jj$ final states must be extracted from the transverse
mass distribution.

\begin{figure}[thb]
\begin{center}
\mbox{
\epsfig{file=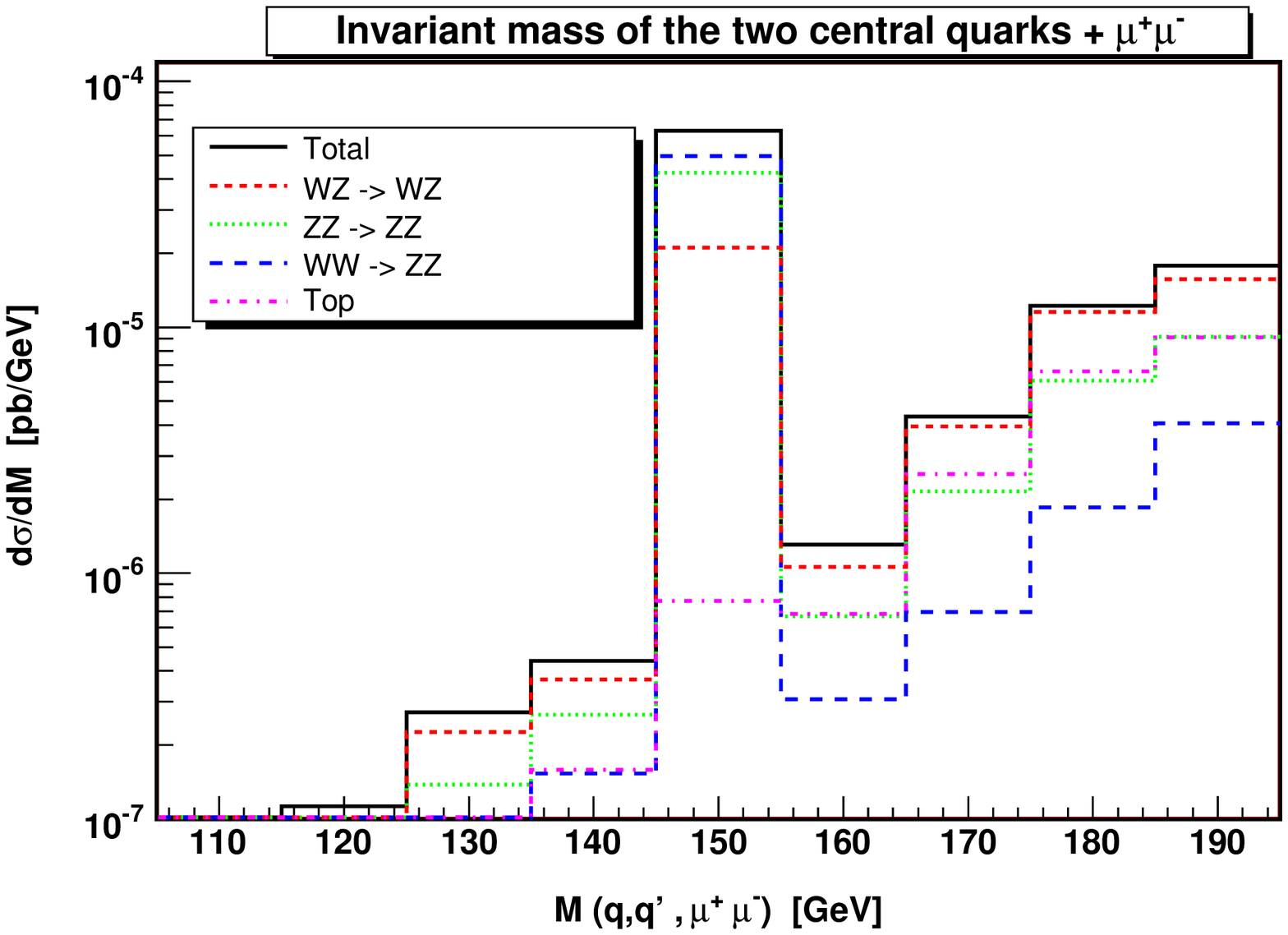,width=8.5cm}
\epsfig{file=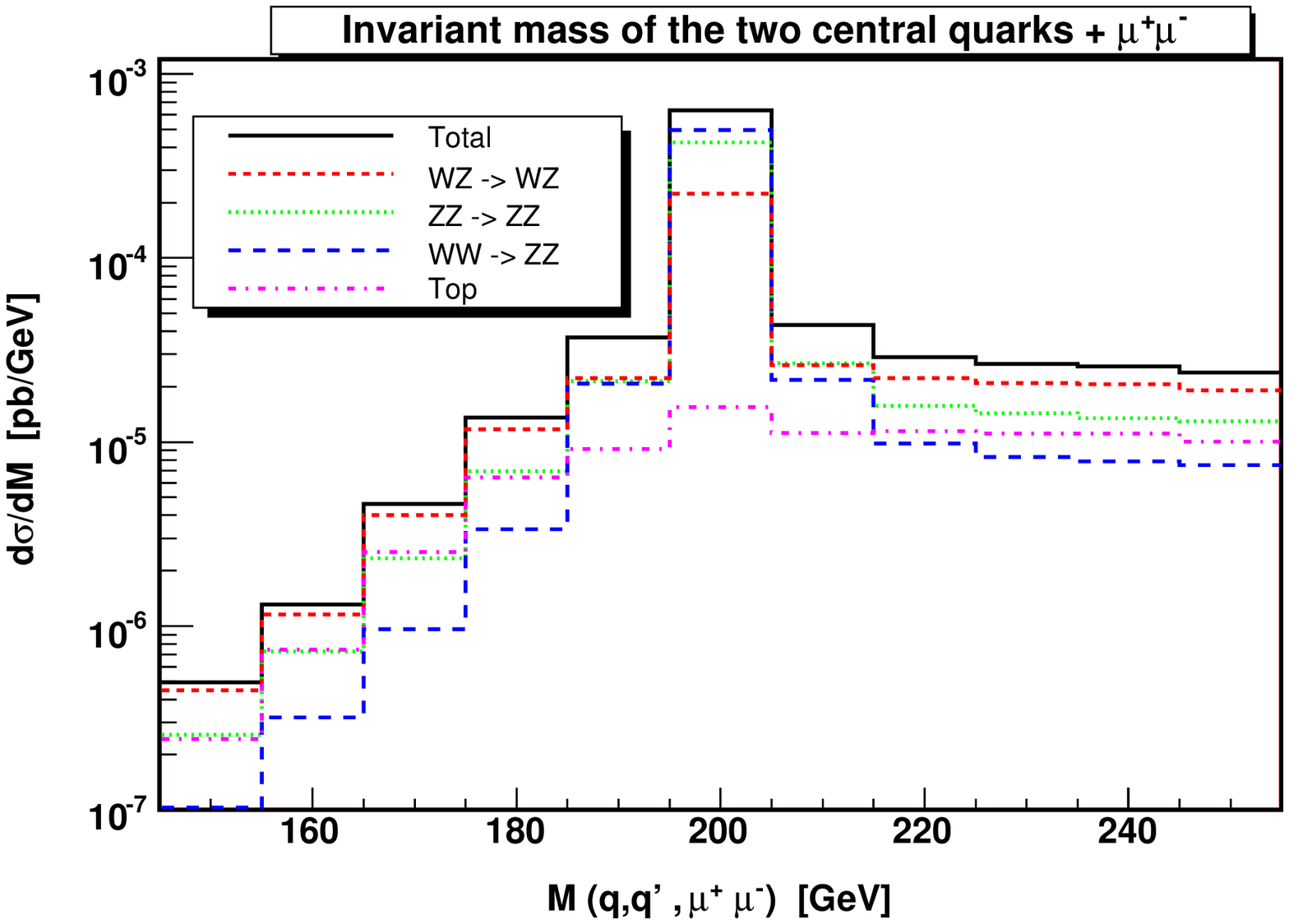,width=8.5cm}}
\caption{ Distribution of the invariant mass $M(VZ)$
of the two candidate vector bosons for a Higgs mass of 150 GeV
and 200 GeV. The contribution of the various subprocesses is evaluated as in
\sect{PhysSub}.}
\label{mh150_200}
\end{center}
\end{figure}

As an illustration of \Phantom capabilities, 
the four body invariant mass distribution of the $\mu^+\mu^-$ pair and
the two most central quarks in $4q\mu^+\mu^-$ final states
in the neighborhood of the Higgs peak is shown
in \fig{mh150_200} for M(H)=150 GeV and M(H)=200 GeV.
The plot on the left is obtained by zooming into the area around the Higgs peak in
\fig{sub-proc}. Both plots show the results for all reactions which contain the
different subprocesses, as described in \sect{PhysSub}, as well as the total
distribution.
Assuming a mass resolution of $\pm 10$ GeV around the peak,
the EW irreducible background amounts to about 3\%(13\%)
for M(H)=150(200) GeV.
An order of magnitude assessment of the statistical significance of such a cross
section for Higgs discovery can be obtained by comparison with the $jjl\nu$
channel.
The main reducible background is QCD $V+nj$ production.
Assuming that the effect of
acceptance cuts are similar in the \WW and in the \ZZ  channel, one can estimate
the ratio of the significancies $S$ in the two cases as

\be
\frac{S(l^+l^-jj)}{S(l\nu jj)} \approx
\frac{\sigma_{qqH}\cdot BR(H\rightarrow ZZ)\cdot BR(ZZ\rightarrow l^+l^-jj)}
     {\sigma_{qqH} \cdot BR(H\rightarrow WW) \cdot BR(WW\rightarrow l \nu jj)}
     \times
\frac{\sqrt{\sigma_{l\nu 4j}}}{\sqrt{\sigma_{l^+l^-4j}}}
\ee

Since  $\sigma_{l\nu 4j}/\sigma_{l^+l^-4j} \approx 10$ \cite{Mangano:2002ea}
and $ BR(ZZ\rightarrow l^+l^-jj)/ BR(WW\rightarrow l \nu jj) \approx 1/3$
we are left with
\be
\frac{S(l^+l^-jj)}{S(l\nu jj)} \approx
\frac{BR(H\rightarrow ZZ)}{BR(H\rightarrow WW)}
\ee

For M(H)$>$200 GeV the ratio of the two branching ratios is about 0.5 and 
on the basis of the studies of the ATLAS \cite{Atlas_HinWW}
and CMS \cite{CMS:Higgs1} collaborations for the \WW channel, one
expects a good statistical significance, of order five, for 
$qqH,\ H\rightarrow ZZ,\ ZZ\rightarrow l^+l^-jj$.
This naive estimate is in rough agreement with the analysis of
\cite{Atlas_HinVBF} which obtains significancies slightly below four in the mass
range $200 < M(H) < 300$ GeV.

Below the $ZZ$ threshold, the Higgs branching ratio to $ZZ$ reaches
about 0.08 at $M(H)\approx 150$ GeV.
For a luminosity of 30 $fb^{-1}$ about 40 
events are expected in the $H\rightarrow l^+l^-jj,\ l=e,\mu$ channel.
Only a complete analysis
including all backgrounds and full detector simulation could tell whether
this is enough for establishing a Higgs signal in this range of masses
in the vector fusion channel.

\begin{figure}[hbt]
  \begin{center}
    \includegraphics[width=12cm]{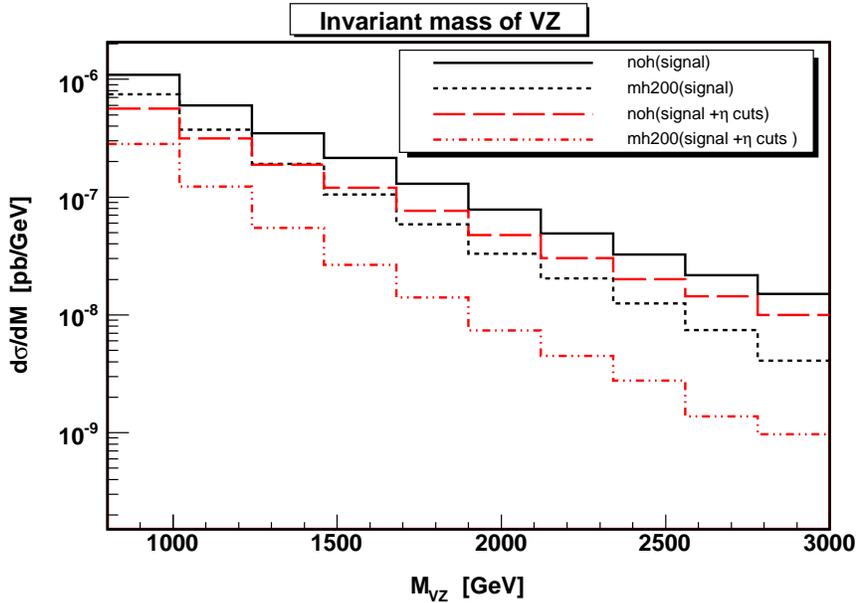}
\caption{Invariant mass distribution  for $M(VZ)>800$ GeV.
The full and long--dashed line refer to the no-Higgs case, 
the short--dashed and dot--dashed ones to M(H)=200 GeV.
All results satisfy the standard acceptance cuts.
The full and short--dashed curves present the
results for our signal definition. For the 
long--dashed and dot--dashed histograms
we have further required $\vert\eta (Z_{ll})\vert<\ $2 and
$\vert\eta (q_V)\vert<\ $2.}
 \label{noh_mh200}
  \end{center}
\end{figure}

\begin{figure}[hbt]
  \begin{center}
    \includegraphics[width=12cm]{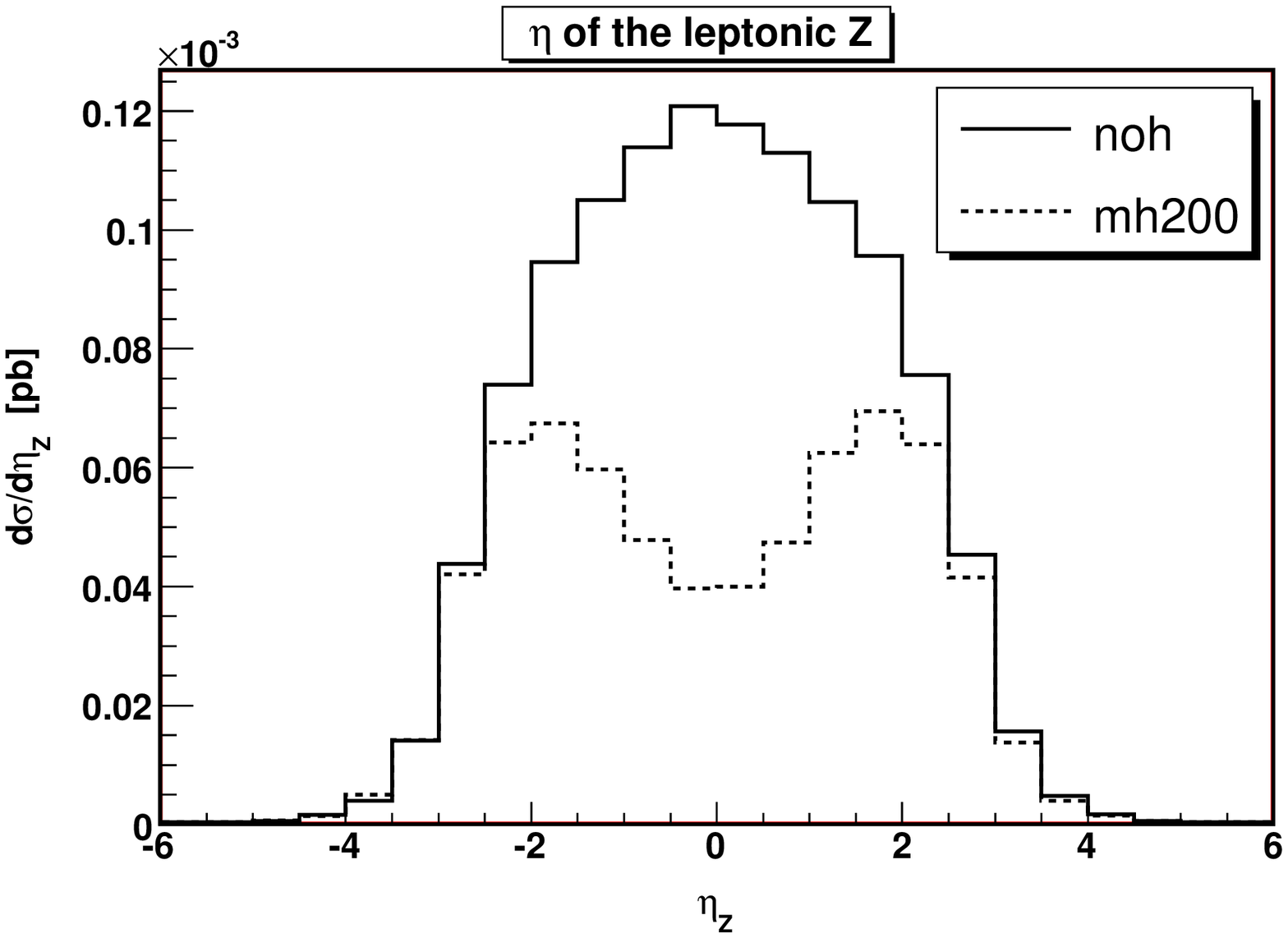}
\caption{Pseudorapidity distribution of the lepton pair
for the no-Higgs case (full)
and for M(H)=200 GeV (dashed). All events satisfy $M(ZV)>800$ GeV.}
 \label{noh_mh200_eta}
  \end{center}
\end{figure}

\subsection{The high VV mass region}
\label{sec:highmass}

In the absence of firm predictions in the strong scattering
regime, trying to gauge the possibilities of discovering signals of
new physics at the LHC requires the somewhat arbitrary definition of a model of
\VVL scattering beyond the boundaries of the SM. Some of these models predict
the formation of spectacular resonances which will be easily detected.
For some other set of parameters in the models only rather small effects are
expected, see for instance \cite{unitarization,butterworth02}.

The simplest approach is to consider the SM in the presence of a very 
heavy Higgs.
While this entails the violation of perturbative unitarity, the linear rise of
the cross section with the invariant mass squared in the hard \VV
scattering will be masked by the decrease of the parton luminosities at large
momentum fractions and, as a consequence, will be particularly challenging to
detect. At the LHC, the offshellness of the incoming vector bosons will further
increase the difference between the expectations based on the behaviour of on
shell \VV scattering and the actual results.  
For $M_H>$10 TeV, all Born diagrams with Higgs propagators become
completely
negligible in the Unitary gauge, and the predictions for all processes in
\eqn{6f} reduce to those in the $M_H \rightarrow \infty$ limit.
In this section we compare this minimalistic definition of
physics beyond the Standard Model, which we call the no Higgs scenario,
with the predictions of the SM with a light Higgs.

An analysis of selection cuts capable to increase the difference between
the no-Higgs and light Higgs cases
could provide some guidance for the search of signals of new physics in boson
boson scattering.

As already mentioned, in the absence of the Higgs, all SM scattering processes
between on shell weak vector bosons grow linearly with the center of mass
energy squared, with the exception of \ZZ \ra \ZZ. 
Therefore all possible
reactions, \ZW \ra \ZW, \WW \ra \ZZ, \ZZ \ra \ZZ, should be carefully
investigated.

An interesting possibility is to investigate whether
there exist or not an elementary Higgs boson by measuring the 
\VV cross section at large M(\VV). 
Previous studies \cite{Accomando:2005hz} 
have shown that kinematical distributions are quite
insensitive to the value of the Higgs mass provided it is much smaller than
the invariant mass of the \VV system.
  
In order to isolate the \VV fusion signal, and more generally two vector boson
production, from all
other six fermion final state processes and investigate EWSB, different kinematical 
cuts have been applied to the simulated events.

First of all, single top production is vetoed as discussed in \sect{PhysSub}.
Second, 
the invariant mass of the two charged leptons has to reconstruct the mass of
a \Z, and  is required to be in the range $M_Z \pm 10$ GeV.
In \VV fusion an additional \W or \Z decaying hadronically
is expected to be present.
Therefore events are required to contain two quarks with the correct flavour
content
to be produced in \W or  \Z decay, with an invariant
mass of $\pm$ 10 GeV around the central value of the appropriate EW boson.
If more than one combination of two quarks satisfies these requirements, the one
closest to the corresponding central mass value is selected. This combination 
will in the following be assumed to originate from the decay of an EW vector
boson.

\begin{table}[tbh]
\begin{center}
\begin{tabular}{|c|c c|c c|c|}
\hline
$M_{cut}$& \multicolumn{2}{c|}{\hspace*{6.5mm}NoHiggs \hspace*{6.5mm}}
& \multicolumn{2}{c|}{M(H)=200 GeV }& Ratio \\
\hline
800 GeV &\hspace*{2.mm}31 & (14,17)&\hspace*{2.mm}12 & (7,5) & 2.59\\
900 GeV &\hspace*{2.mm}25 & (12,13)&\hspace*{2.mm}8 & (5,3)& 3.12\\
1.0 TeV &\hspace*{2.mm}19 & (9,10)&\hspace*{2.mm}6 & (4,2)& 3.16\\
1.1 TeV &\hspace*{2.mm}16 & (7,9)&\hspace*{2.mm}5 & (3,2)& 3.20\\
1.2 TeV &\hspace*{2.mm}13 & (6,7)&\hspace*{2.mm}3 & (2,1)& 4.33\\
1.3 TeV &\hspace*{2.mm}11 & (5,6)&\hspace*{2.mm}2 & (1,1)& 5.50\\
1.4 TeV &\hspace*{2.mm}9 & (4,5)&\hspace*{2.mm}2 & (1,1)& 4.50\\
\hline
\hline
\multicolumn{6}{|c|}{$\Delta R = 0.4$} \\
\hline
$M_{cut}$& \multicolumn{2}{c|}{\hspace*{6.5mm}NoHiggs \hspace*{6.5mm}}
& \multicolumn{2}{c|}{M(H)=200 GeV }& Ratio \\
\hline
800 GeV &\hspace*{2.mm}18 & (8,10)&\hspace*{2.mm}10 & (6,4) & 1.80\\
900 GeV &\hspace*{2.mm}12 & (5,7)&\hspace*{2.mm}6 & (4,2)& 2.00\\
1.0 TeV &\hspace*{2.mm}8 & (4,4)&\hspace*{2.mm}4 & (2,2)& 2.00\\
1.1 TeV &\hspace*{2.mm}5 & (2,3)&\hspace*{2.mm}3 & (2,1)& 1.60\\
\hline
\hline
\multicolumn{6}{|c|}{$\Delta R = 0.5$} \\
\hline
$M_{cut}$& \multicolumn{2}{c|}{\hspace*{6.5mm}NoHiggs \hspace*{6.5mm}}
& \multicolumn{2}{c|}{M(H)=200 GeV }& Ratio \\
\hline
800 GeV &\hspace*{2.mm}12 & (5,7)&\hspace*{2.mm}8 & (5,3) & 1.50\\
900 GeV &\hspace*{2.mm}8 & (4,4)&\hspace*{2.mm}5 & (3,2)& 1.60\\
1.0 TeV &\hspace*{2.mm}5 & (2,3)&\hspace*{2.mm}3 & (2,1)& 1.60\\
\hline
  
\end{tabular}
\end{center}
\caption{Number of events as a function of the minumum
invariant mass of the $ZV \rightarrow \mu^+\mu^-jj$ pair for L=100 $fb^{-1}$.
All events satisfy
$\vert\eta (Z_{ll})\vert<\ $ 2 and $\vert\eta (q_V)\vert\ <$ 2.
In brackets we show the contribution of the (\ZW,\ZZ) final states.}
\label{events}
\end{table}

In a third step, in order to reject events which can be identified with the
production of three vector bosons, the flavour content and the invariant mass 
of the two remaining quarks is compared with a \W and a \Z. If compatible 
within 10 GeV with either, the event is rejected.
The events satisfying all these constraints will constitute the ``signal'' sample.

\begin{figure}[tbh]
\begin{center}
\mbox{
\epsfig{file=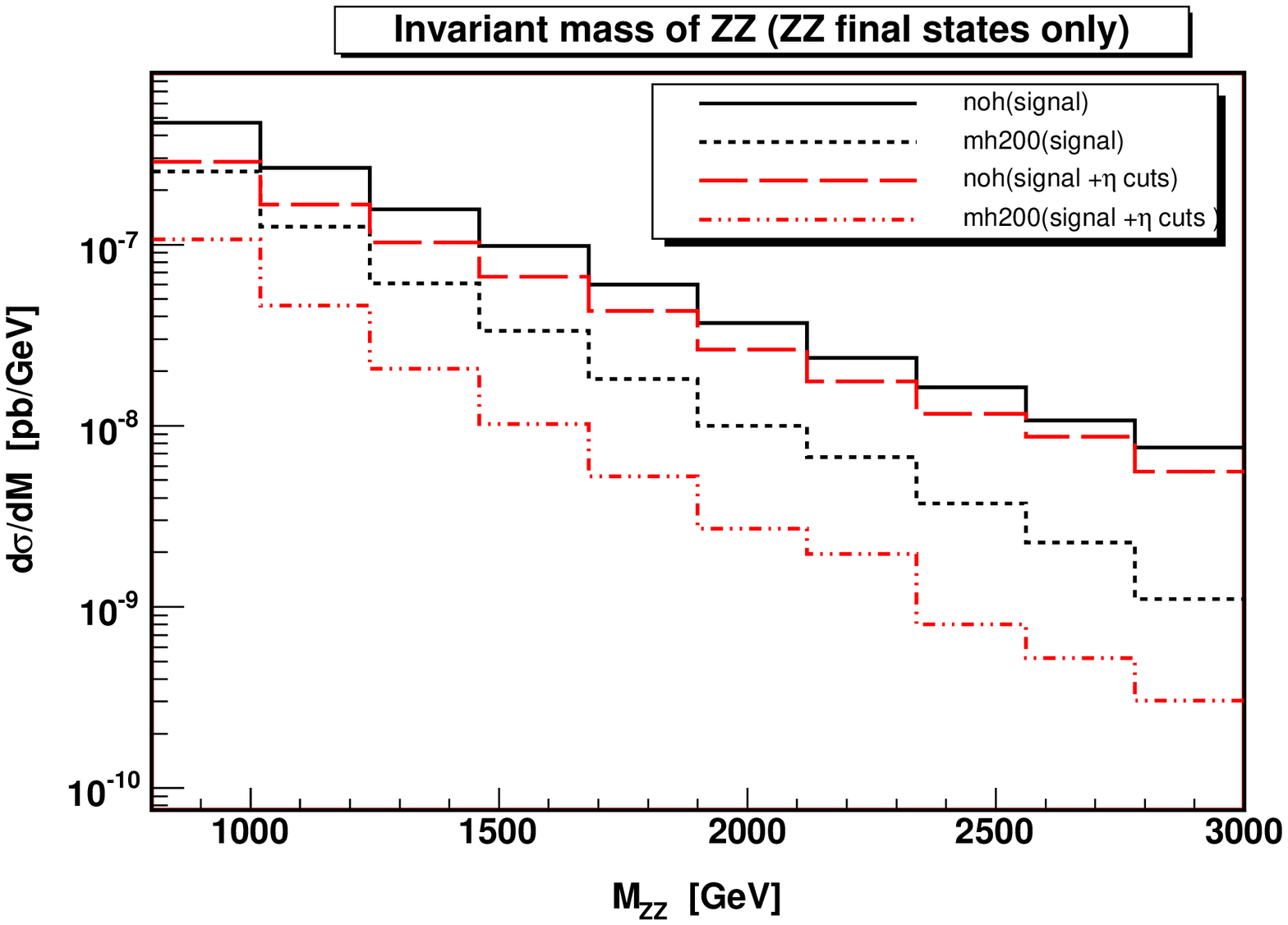,width=8.5cm}
\epsfig{file=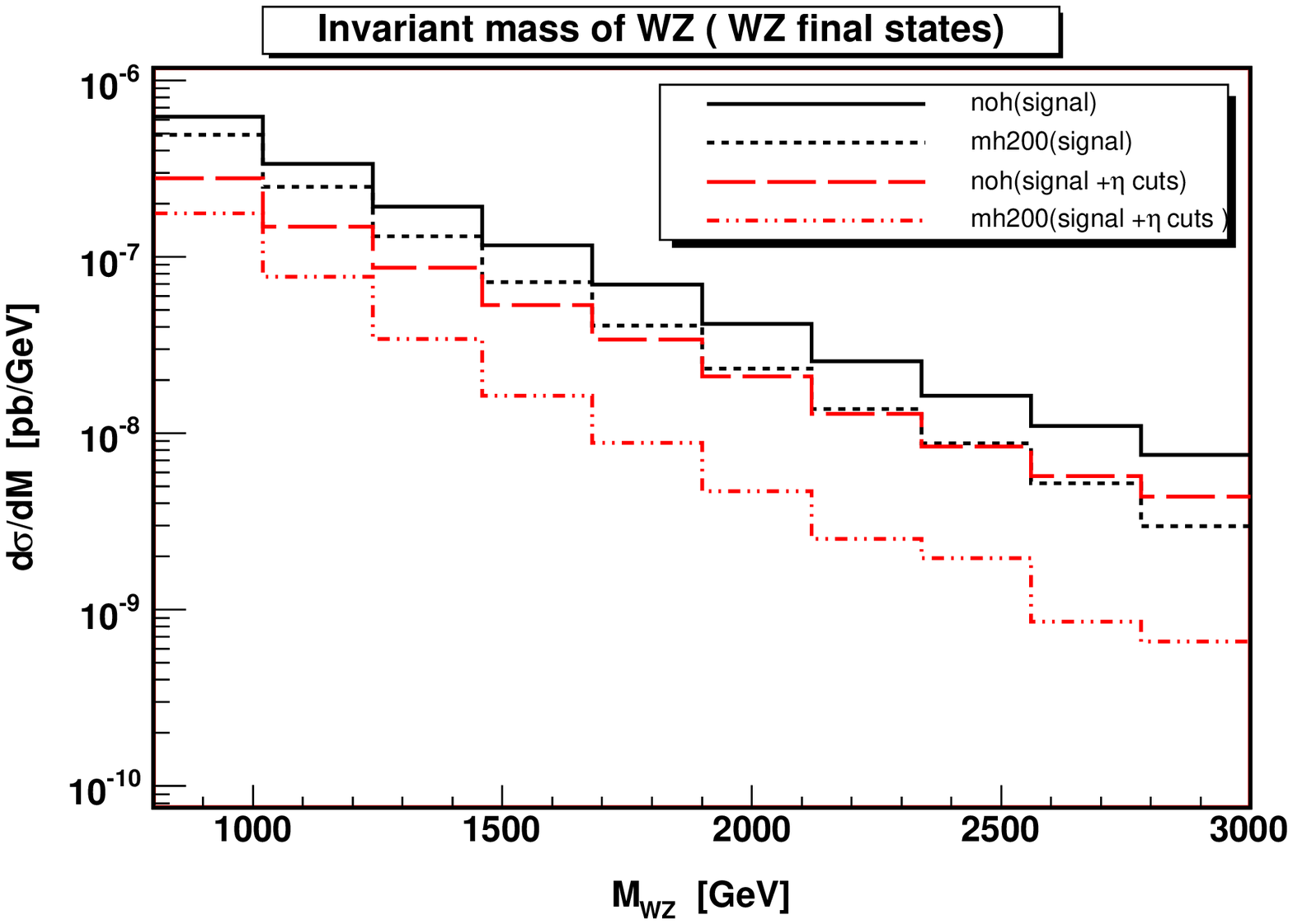,width=8.5cm}
}
\caption{ $M(ZZ)$ (left) and $M(WZ)$ (right) invariant mass distribution  for
$M(VZ)>800$ GeV.
The full and long--dashed lines refer to the no-Higgs case,
the short--dashed and dot--dashed one to M(H)=200 GeV.
The full and short--dashed curves present the
results for our signal definition. For the 
long--dashed dot--dashed histograms
we have further required $\vert\eta (Z_{ll})\vert<\ $2 and
$\vert\eta (q_V)\vert<\ $2.}
\label{noh_mh200_ZZ_WZ}
\end{center}
\end{figure}

In \fig{noh_mh200} we present the invariant mass distribution of the two charged
leptons and the two jets associated with the vector boson decay for
M(H)=200 GeV and for the no-Higgs case.
A number of selection cuts have been studied in order to increase the difference
between the two Higgs hypotheses. Simple requirements of centrality of the
lepton pair and of the candidate second vector boson have proved to be the most
effective. The pseudorapidity distribution of the charged lepton pair 
in the two cases is shown in \fig{noh_mh200_eta}.
The long--dashed dot--dashed distributions in \fig{noh_mh200}
have been obtained with
the additional contraints that  $\vert\eta (Z_{ll})\vert<\ $ 2 and
$\vert\eta (q_V)\vert\ <$ 2, where $q_V$ refers to the quarks which are
associated with the vector boson decay.
The corresponding distributions for the \ZW and \ZZ final states are presented
in \fig{noh_mh200_ZZ_WZ}. The cross section for $qqZW$ and $qqZZ$ production
are similar, however the discrepancy between the no-Higgs case
and the M(H)=200 GeV  is larger for the $qqZZ$ final state.

In the upper part of \tbn{events} we present the number of events as a function of the
minimum invariant mass of the $\mu^+\mu^-jj$ system for L=100 $fb^{-1}$ with the
set of cuts shown in \tbn{standard-cuts}. 
In brackets we also give the separate results for the \ZW and \ZZ final states.
The number of
events is smaller than the expected yield in the $4q\mu\nu$ channel
\cite{Accomando:2005hz} but the differences between the two Higgs hyphotheses
are larger. In fact, similar ratios are obtained with
comparable number of events. 

In \tbn{events} we also show the effect of requiring a minimum
$\Delta R$ separation
among colored partons. The expected number of events decreases sharply, by about
a factor of two for $\Delta R = 0.4$ and by about a factor of three
for $\Delta R = 0.5$, in the NoHiggs case. The statistics for a light Higgs  is
less affected because the vector boson distribution is less central in this case
and it is precisely the jets originating from the most central and higher $p_T$
\W's and \Z's which are most likely to merge into one jet under the effect of a 
$\Delta R$ cut.
As already mentioned in \sect{PhysSub}, this issue is well known
and various approaches have been tried in the literature.
ATLAS \cite{ATLAS-TDR} (Sec. 9.3.1.3, 19.2.10.2),
favors using a small cone  $\Delta R = 0.2$ for the
determination of the jet centroid and then a larger cone $\Delta R = 0.4$ for
collecting the energy flow of the jets. In QCD studies at the LHC a typical
separation $\Delta R = 0.5$ is adopted. 
In Ref. \cite{butterworth02,butterworth07} a different approach has been
proposed. First, jets with a total invariant mass in the neighborhood of the
electroweak vector meson mass are selected. Then these jets are forced to
divide into two subjets, by varying the separation parameter $y$ in the
$k_\perp$ scheme.
For jets originating from a vector meson decay the
scale $\sqrt{yp^2_T}$ at which the heavy jet splits into two subjets is
typically close to the
vector mass, while for standard QCD jets the splitting scale is much smaller.
The subject of jet reconstruction algorithms is still lively debated.
Since EW vector bosons are crucial to many investigations at the
LHC, we expect that a scheme which allows to distinguish jets produced in
the decay of high $p_T$ \W's and \Z's will be devised for this kind of
specialized studies.

\begin{figure}[thb]
\begin{center}
\mbox{
\epsfig{file=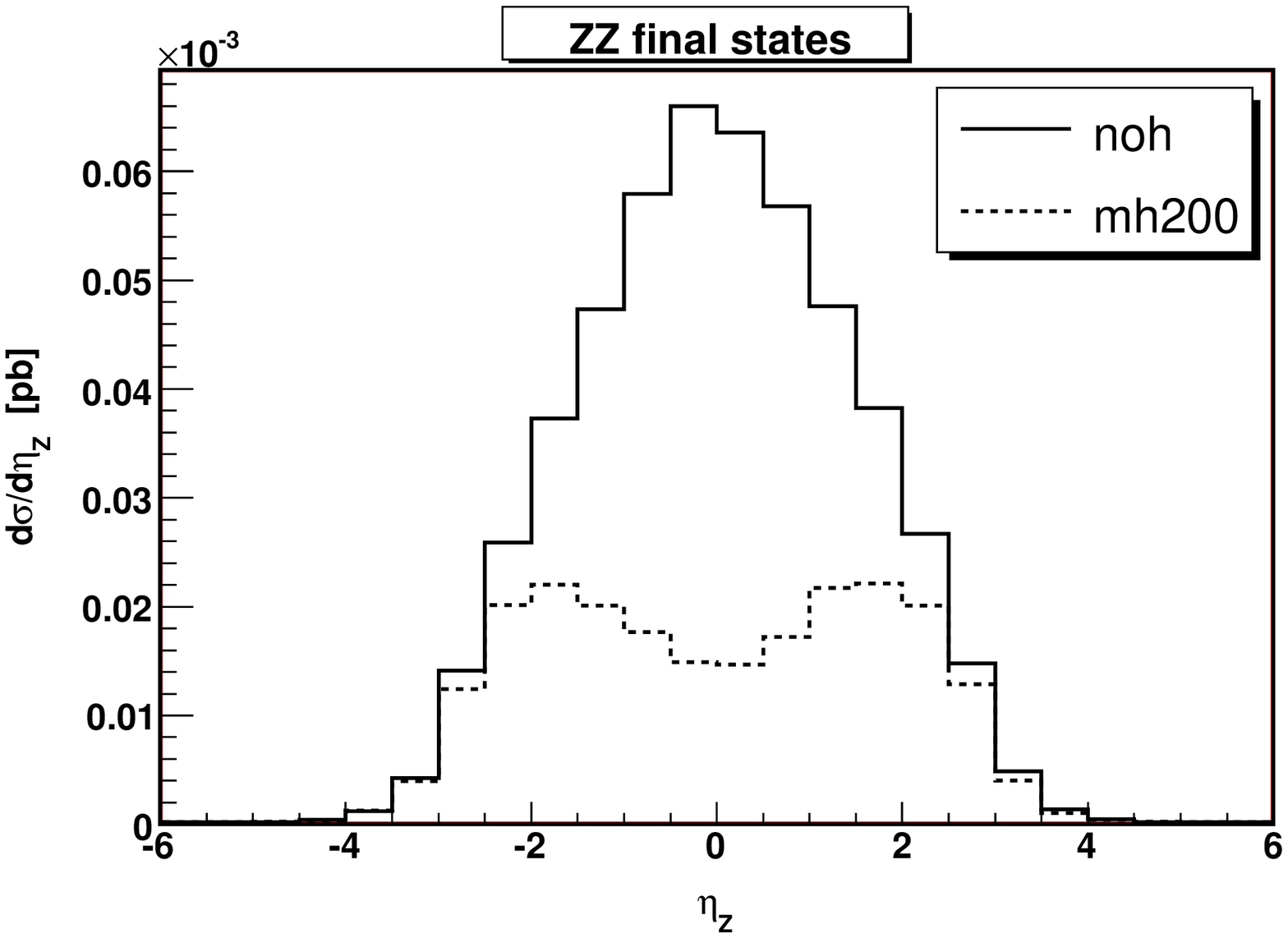,width=8cm}
\epsfig{file=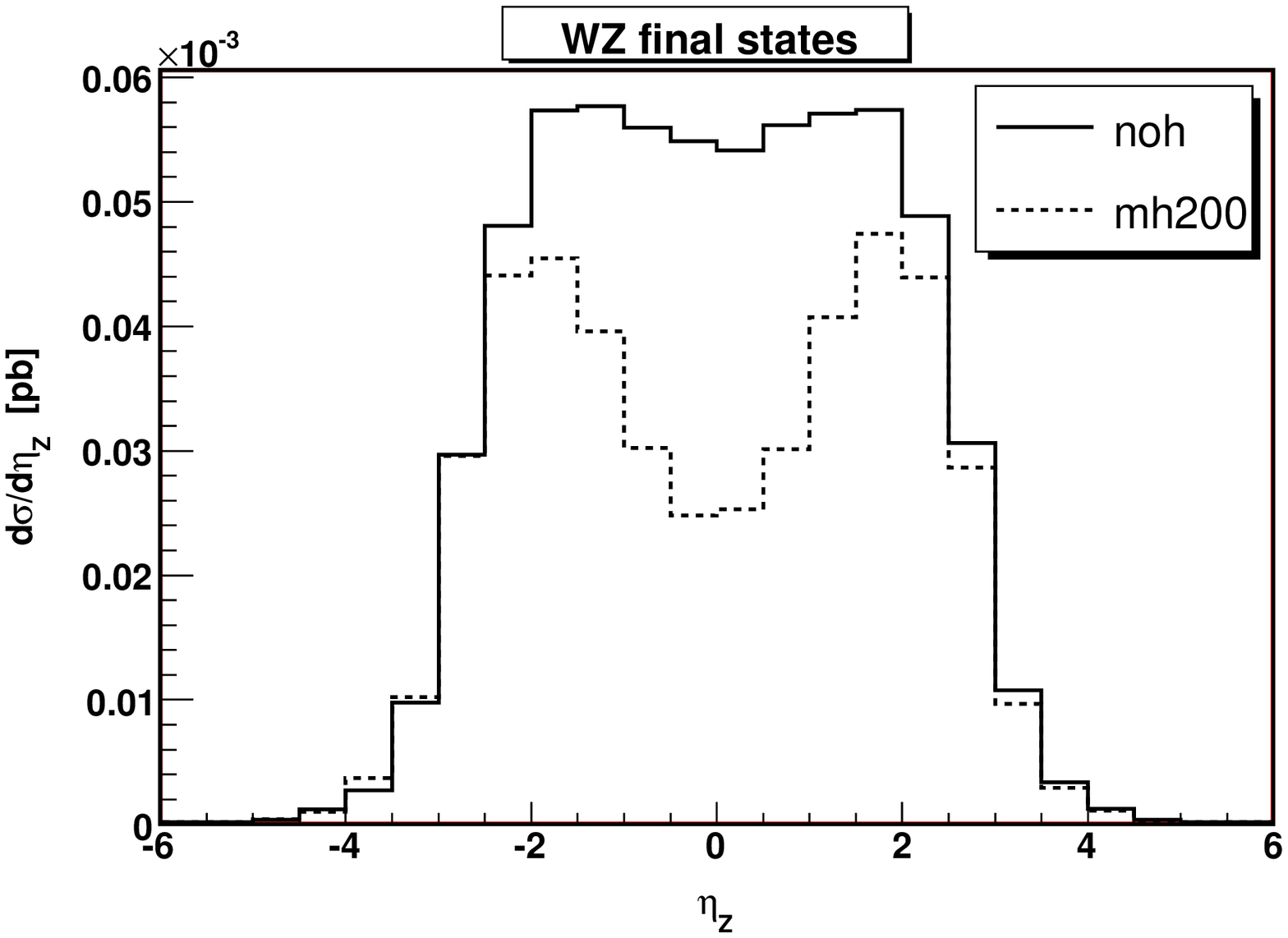,width=8cm}
}
\caption{Pseudorapidity distribution of the lepton pair in $qqZZ$
(left) and $qqZW$ (right) final states.
The full line refers to the no-Higgs case, the dashed one to M(H)=200 GeV.
In all cases $M(VZ)>800$ GeV.}
\label{ZZ_WZ}
\end{center}
\end{figure}

At the LHC, the expected mass resolution for vector bosons decaying to dijets
is about $5\div10$ GeV, depending on the boson transverse momentum
\cite{ATLAS-TDR}. This
makes it quite difficult to separate $ZZ \rightarrow \mu^+\mu^- jj$ from
$ZW \rightarrow \mu^+\mu^- jj$ final states on the basis of the invariant mass
of the jet pair. It is therefore of interest to explore alternative means of
separating the two final states. The low energy theorem
\cite{LowEnergyTheorem} predicts that $A(W^+W^- \rightarrow ZZ)= s/v^2$ where 
$A$ is the scattering amplitude, $s$
is the usual Mandelstam variable and $v$ is the coupling strength of the gauge
current to the Goldstone boson, which in the SM coincides with the vacuum
expectation value of the Higgs field. As a
consequence of crossing symmetry then $A(WZ \rightarrow WZ)= t/v^2$.
Therefore, averaging between the two orientation of the incoming \ZW system,
and neglecting vector boson masses, one expects, in the reaction center of mass,
$d\sigma/d\cos\theta \sim {\mathrm const}$ for  $W^+W^- \rightarrow ZZ$ and
$d\sigma/d\cos\theta \sim (1+\cos\theta^2)$ for $WZ \rightarrow WZ$
where $\theta$ is the scattering angle in the boson boson
center of mass.
This has led us to study the pseudorapidity distribution of the lepton pair
in $qqZZ$ and $qqZW$ final states separately, as shown in \fig{ZZ_WZ}.
Despite the fact that we are not in the center of mass of the $VZ$ system, that
the incoming vector bosons are not on their mass shell and that we are plotting
the distribution of an angular variable which is not the cosine of the center of
mass scattering angle, the general prediction that the two final states have
different distributions is verified. 
In the NoHiggs case, the \ZW  channel distribution is almost flat in the region
$\vert \eta \vert <2$ with small peaks in the forward and backward
direction, as suggested by the LET, while the \ZZ one peaks at zero
rapidity.
It is somewhat surprising, but quite welcome, that  the two distributions are
markedly different also when a light Higgs boson is present in the spectrum
as expected in the SM.
It should be mentioned that the angular distribution depends on
the vector boson pair invariant
mass. The plot for $M(VZ)>300$ GeV, which we do not show, displays a similar,
rather central, behaviour for the two processes.

\begin{figure}[hbt]
  \begin{center}
    \includegraphics[width=12cm]{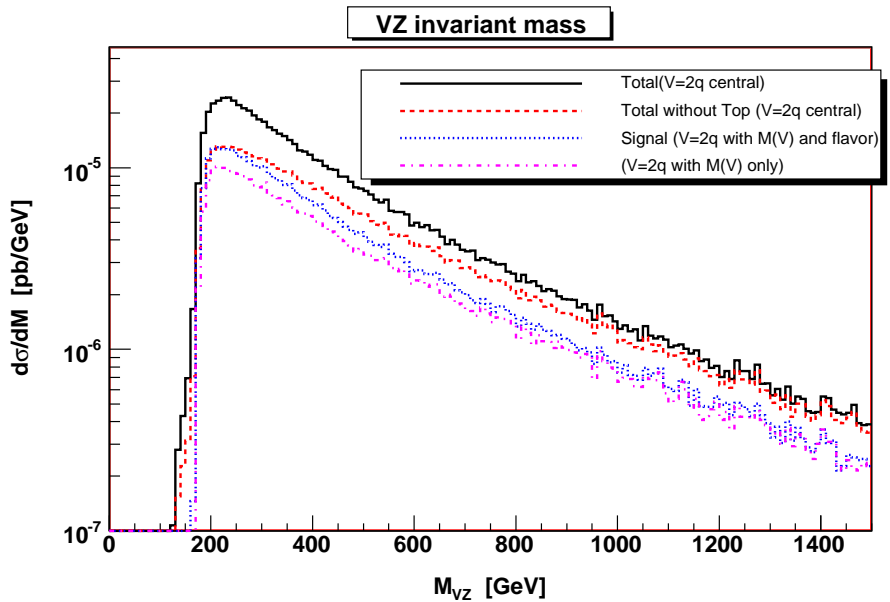}
\caption{Invariant mass distribution of the lepton pair and the two jets from
boson decay for the no-Higgs case.
The solid(dashed) line is obtained identifying the two most central jets as the vector boson decay products
before(after) top vetoing.
The dotted line is obtained requiring the correct flavour content for the jets
identified as decay products of both the vector boson and the top.
The dot--dashed lines is produced using solely invariant mass informations
to identify the vector boson and the top decay products.}
 \label{tot_signal_exp}
  \end{center}
\end{figure}

The selection procedure employed for \fig{noh_mh200} and \tbn{events}
is not fully realistic: no flavour information will be
available for light quarks and $b$'s will be tagged only in the central part of
the detector. At this stage we want to isolate as much as
possible the \VV fusion signal from all other production channels,
with the aim to identify a possible signal definition which could play the role
which was played by {\tt CC03} at LEP2 \cite{ref:CC03}, capturing the essence of the
physical phenomena under investigation and allowing comparisons between the
results from different collaborations. 
It is however of interest to investigate whether the relevant distributions
are sensitive to the details of the selection cuts.
In \fig{tot_signal_exp} we compare the invariant mass distribution
of the two charged leptons and the two quarks associated with the vector
boson obtained with two different methods.
In the first case (dotted line) we select only quark pairs that have the correct flavour
content
to be produced in \W or  \Z decay while in the second (dot--dashed line)
we consider all quark
pairs. In the more realistic setting the top veto is applied to any triplet
of quarks with a total  invariant mass between 160 and 190
GeV; moreover alla events in which two quark pairs with mass between
$M_W-10$ GeV and $M_Z+10$ GeV are present are discarded.
In both cases we identify the candidate vector boson with the quark pair
whose mass is closest to the nominal vector boson mass.
The two distributions differ by about 20\% at small invariant masses
but agree quite nicely at invariant masses above 800 GeV,
showing that our results based on Monte Carlo level flavour information
are not seriously degraded when selection procedures closer to the actual
experimental practice are adopted.
For comparison we also present the invariant mass distribution obtained by
identifying the two most central jets as the vector boson decay products
before (solid line) and after (dotted line) top vetoing.
 
\section{Conclusions}

In this paper we have studied all $q_1 q_2 \rightarrow q_3 q_4 q_5 q_6 l^+l^-$
processes at order {$\O(\alpha_{em}^6)$} at the LHC
using for the first time a full fledged six fermion Monte Carlo event generator.
We have studied Higgs production and its irreducible EW background
in vector boson fusion followed by the decay
chain $H\rightarrow ZZ\rightarrow l^+l^-jj$,
including exactly all electroweak irreducible backgrounds and in particular
the interference of EW $ZZ+2j$ and $ZW+2j$ production with the signal.
The EW irreducible background in the neighborhood of the Higgs peak 
amounts to about 1.5\%(6\%) for M(H)=150(200) GeV.
We have examined how simple kinematical cuts can be applied at generator level
to extract the \VV signal from the irreducible background.
In the high mass region we have compared the case of a relatively light Higgs
with the no-Higgs case. The integrated cross section for the latter is
more than twice that in the former for a minimum
invariant mass of the $ZV$ pair of about 800 GeV. Summing up the muon and the
electron channels, about 25 events are expected in the light Higgs case
for L=100 $fb^{-1}$.
These results are encouraging and show that a more complete analysis, including
all QCD backgrounds and an accurate study of jet separation algorithms in the
high invariant mass region, is worthwhile.

\end{document}